 \documentclass[
twocolumn,showpacs, amsmath, amssymb, floatfix,
nofootinbib,wrapfloat byrevtex]{revtex4}

 \usepackage{amsmath,mathrsfs,graphicx,epstopdf,placeins,float}
 \usepackage{booktabs}
 \usepackage{multirow}
\usepackage{color}

 \newcommand{\beq}[1]{\begin{equation}\label{#1}}
 \newcommand{\eeq}{\end{equation}}
 \newcommand{\bea}[1]{\begin{eqnarray}\label{#1}}
 \newcommand{\eea}{\end{eqnarray}}
 \newcommand\figcaption{\def\@captype{figure}\caption}
 \newcommand\tabcaption{\def\@captype{table}\caption}

\begin{document}

 \title{Geometric Description of Schr\"{o}dinger Equation in Finsler and Funk Geometry}
\author{Asma Bashir$^{1}$}
\email{kbasmabashir807@gmail.com}
\author{B. Koch$^{2}$}
\email{bkoch@fis.puc.cl}
\author{M. A. Wasay$^{3}$}
\email{wasay31@gmail.com}
\affiliation{$^1$Department of Physics, Government College Women University, Faisalabad 38000, Pakistan.\\
$^2$
Instituto de F\'{i}sica, Pontificia Universidad Cat\'{o}lica de Chile,
Av. Vicu\~{n}a Mackenna 4860,
782-0436 Macul, Santiago, Chile.\\$^3$Department of Physics, University of Agriculture,
Faisalabad 38040, Pakistan.}
 \begin{abstract}
For a system of $n$ non-relativistic spinless bosons, we show by using a set of suitable matching conditions that
the quantum equations in the pilot-wave limit can be translated into a geometric language for a Finslerian manifold.
We further link these equations to Euclidean timelike relative Funk geometry
and show that the two different metrics in both of these geometric frameworks lead to the same coupling.
  \end{abstract}

\maketitle
 \smallskip
\section{introduction}
Geometric theories, in particular the link between these theories and quantum mechanics is
a very active field of research.
Spacetimes equipped with Lorentzian metric have been serving as geometrical framework to discuss the dynamics of the particles for over hundred years. However, \cite{1,2} suggest that modifying gravity at times could be helpful for a more comprehensive depiction of nature. Several modifications of GR have been discussed in \cite{4,5,6}. In particular, Finsler geometry is a natural generalization of the Lorentzian metric geometry. There are several evidences supporting this idea because this extension of GR maintains a simultaneous geometrization of gravity, observer and causality, e.g., see \cite{32} and references therein.

There are also several generalizations of the Finsler geometry, i.e., Euclidean timelike relative Funk geometry, hyperbolic timelike Funk geometry, timelike relative spherical Funk geometry etc. These theories of timelike spaces were introduced in \cite{7}. It was shown in \cite{8} that these timelike spaces are merely timelike Finsler spaces. This approach entails that the metric of such a space is determined by timelike norm on each tangent space in such a way that the distance between two points is infimum of length of path joining them and the length is defined between events using a timelike distance function.

On the other hand, quantum mechanics is one of the most profound and successful (algebraic and linear) mathematical models which helps describe the physical reality around us. However, the quantum phenomenon are obscure and unpredictable because of their probabilistic nature. Classical mechanics however is predictable, geometric, and nonlinear. Geometric in the sense that phenomenon described by classical mechanics is usually defined over a symplectic manifold. The quest to unify these two seemingly different approaches to reality is thus natural and is open for a long time and there has been a lot of effort in this direction.

In order to reformulate quantum mechanics in geometrical terms, one needs to associate physical reality to
objects and to define the background space. The well-known Copenhagen interpretation suggests that one can associate physical reality with a particle only after the measurement, before the measurement it is meaningless to talk about physical properties of a particle due to the wavefunction collapse upon measurement.

Although different interpretations provide different groundwork to study quantum mechanics, a particularly nice approach
is the pilot-wave theory, which provides an extensive framework in order to perceive the theory conceptually and completely.
The idea of merging quantum and geometry to provide a unified framework came into play few years after the development of pilot-wave theory and is still an ongoing issue.
In order to associate physical reality to a set of particles, one needs to associate a pilot wave to each particle that guides the specific particle along its trajectory according to the guiding equation
$m_i\frac{dx_i}{dt}=\nabla_i S$.

A quantum phenomenon could be perfectly choreographed with a well chosen hidden variable
approach and could subsequently be reformulated in geometrical way, refs.\cite{9}-\cite{23} represent
such efforts for a geometrical rewriting of quantum laws. A slightly different approach linking quantum mechanics with topological properties, has also been worked on extensively, and a few of these could be found in refs.\cite{top1,top2,top3,top4,top5,top6}.
 The geometrisation of the Klein-Gordon equation for a conformally curved spacetime was done in \cite{23}, and in \cite{29} it was generalized for the nonrelativistic case, i.e., the Schr\"{o}dinger equation. A Finslerian version of the Klein-Gordon equation was presented in \cite{34} and in another relevant work two particle entanglement was written geometrically for Finsler spacetimes \cite{30}. Details about Finsler space can be found in \cite{35,36} and references therein.

The purpose of this paper is to translate the quantum equations in the pilot-wave limit to a geometric language of both Finsler and timelike relative Funk spacetimes in the context of modified gravity (by adding  extra dimensions in the configuration space).

This paper is organized as follows:
Sec.II, describes the physical model and establishes the quantum equations in the pilot-wave limit.
In sec.III, we present the geometric interpretation of quantum equations over Finsler spacetime.
In sec.IV, we present the geometric translation of these equations over timelike relative Funk space.
Sec.V is the summary and conclusion.

\section{\textbf{The Model}}
We will study a system of $n$-spinless bosons in the  non-relativistic limit. Further, we assume that these bosons are free of any external potential in such a way that each particle can freely move along any of the three spatial dimensions. The wave equation for spin-0 bosons in the non-relativistic limit reduces to the Schr\"{o}dinger equation of motion \cite{29}.

The Schr\"{o}dinger equation for multi-particle system could be written as
\bea{}
\left(\sum\limits_{a=1}^n\frac{\hbar}{2M_a}\partial_a^b\partial_{ab}+i\partial_0\right)\psi(x_1,x_2,...,x_n,t_1)=0
\label{third}
\eea
where $\partial_0=\frac{\partial}{\partial t_1}$. In the above equation $M_a$ represents the $n$-particle mass. The indices $b$ and $a$ are used to label the $3$-D flat space and the particle (such that it specifies which one out of the $n$-particle is affected) respectively.

We are interested in pilot-wave theory that interprets not only the dynamics of the particles in a system but also accounts for their configuration. To switch from standard quantum interpretation of Eq.\eqref{third} into the pilot-wave limit, the wave function is factored into amplitude and phase \cite{23,27,28,29,30}, as follows

\bea{}
\psi=Re^{iS/\hbar}\label{0}
\eea

We are concerned with a system of free bosons in classical frame of reference, all sharing a universal time coordinate $t_1$. The quantum phase then reads

\bea{}
S(t_a, \vec{x}_a)=-Mt_1+\tilde S(t_1,\vec{x}_a)
\label{1}
\eea

The same is true for Hamilton's principle function $S_H$, and the amplitude $R$ is given by
\bea{}
R(t_a,\vec{x}_a)=R(t_1,\vec{x}_a)\label{2}
\eea
 One can then rewrite the wavefunction in Eq.\eqref{0} as
\bea{}
\psi(t_1,\vec{x}_a)=R(t_1,\vec{x}_a)e^{i\left(\tilde S(t_1,\vec{x}_a)-Mt_1\right)/\hbar}.
\label{four}
\eea
\subsection{First Equation}
In pilot-wave limit, spin-$0$ bosons evolve in space and time according to
\bea{}
\left(\sum\limits_{a=1}^n\frac{\hbar}{2M_a}\partial_a^b\partial_{ab}+i\partial_0\right)R(t_1,\vec{x}_a)e^{i\left(\tilde S(t_1,\vec{x}_a)-Mt_1\right)/\hbar}=0
\label{bb}
\eea

Here $\partial R/\partial t_1=0$ as for large $t_1$ the amplitude on the average is zero.
Using Taylor expansion in Eq.\eqref{bb} and ignoring higher order terms, we can get

\bea{}
Q=\sum\limits_{a=1}^n\frac{(\partial_a^b\tilde S)(\partial_{ab}\tilde S)}{2M_a}+\dot{\tilde S}-M,
\label{five}
\eea

Eq.\eqref{five} aligns nicely with the classical Hamilton-Jacobi equation and represents the motion of bosons as a moving pilot-wave. The term $Q$ here represents the quantum potential $Q(x_1,x_2,...,x_n,t_1)$ and $R(x_1,x_2,...,x_n,t_1)$ is the amplitude of the associated pilot wave that guides the particle along its trajectory. This is the first non-relativistic quantum equation in pilot-wave interpretation.
\subsection{Second Equation}

For a multi-particle system conservation of probability in position space is found to be
\bea{}
\partial_0(\psi^*\psi)-\sum\limits_{a=1}^n\partial_a^b\left(\frac{i\hbar}{2M_a}(\psi^\ast\overleftrightarrow{\partial}_{ab}\psi)\right)=0\nonumber
\eea
The conserved current with the definition in Eq.\eqref{four} yields the following equation

\bea{}
\partial_0(R^2)+\sum\limits_{a=1}^n\partial_{ab}\left(\frac{R^2}{M_a}(\partial_a^b\tilde S)\right)=0
\label{six}
\eea
Eq.\eqref{six} is the second dBB equation. 

\subsection{Third Equation}

Guided by the following equation

\bea{}
p_a^b=M_a\frac{dx_a^b}{ds}=\partial_a^b\tilde S
\label{seven}
\eea

these particles always have well defined trajectories in terms of their initial positions. One can specify the particle trajectory once the particle location at a particular instant of time is determined. This is the idea behind this framework, it treats that particular position as a hidden variable that gives insight into the physical reality of the system.
This is third dBB equation.
\subsection{Fourth Equation}
The trajectory equation of motion for a system of $n$ non-relativistic  bosonic particles is given by (see appendix B)

\bea{}
\frac{d^2x_a^b}{ds^2}=\sum\limits_{c=1}^n\frac{(\partial_c^d\tilde S)(\partial_a^b\partial_{cd}\tilde S)}{M_a^2}
\label{eight}
\eea

Here, $ds=dt_1$ is the common time dimension for all bosons. Eq.\eqref{eight} could serve as Newton's acceleration law in the pilot-wave limit. This equation is a clear evidence of the essential feature (non-locality) of the pilot-wave interpretation. The term left of the Eq.\eqref{eight} simply represents the acceleration of a specific particle which gets affected by a tidal force $\sum\limits_{c=1}^n\frac{(\partial_c^d\tilde S)(\partial_a^b\partial_{cd}\tilde S)}{M_a}$ created by all bosons in the system. That is to say, the entire system is responsible for a particular position and hence for trajectory followed by each particle.
This is analogous to the classical picture but different in the sense that as the particles get far apart (in classical theory) the strength of non-local influence decreases. However in hidden variable approach the non-local influence might be strengthened at large distance.

Eqs.\eqref{five}-\eqref{eight} is a set of four non-relativistic equations for the case of multi-particle system in the pilot-wave interpretation. The parameters $R$ (the amplitude of pilot wave), the quantum phase $\tilde{S}$, and quantum potential $Q$ in these equations depend on $1+3n$ coordinates, $3n$ of space and $1$ of time.

In single index notation, the coordinates can be specified as

\bea{}
[t_1,x^A]=[t_1,(\vec{x}_1^1,\vec{x}_1^2,\vec{x}_1^3,...,\vec{x}_n^1,\vec{x}_n^2,\vec{x}_n^3)] \nonumber
\eea
such that $\partial_a^b\rightarrow\partial^A$ and $\partial_{ab}\rightarrow\partial_A$; with
\bea{}
Q=\frac{\hbar^2}{2M_a}\frac{\partial^A\partial_AR(t_1,\vec{x_a})}{R(t_1,\vec{x_a})}.\nonumber
\eea


The set of four equations, obtained by manipulating non-relativistic wave equation in the context of hidden variable theory, can now be written as
\bea{}
Q=\frac{(\partial^A\tilde S)(\partial_A\tilde S)}{2M_a}+\dot{\tilde S}-M,
\label{nine}
\eea
\bea{}
\partial_A\left(\frac{R^2(\partial^A\tilde S)}{M_a}\right)+\partial_0(R^2)=0,
\label{ten}
\eea
\bea{}
p^A=M_a\frac{dx^A}{ds}=\partial^A\tilde S,
\label{eleven}
\eea
\bea{}
\frac{d^2x^A}{ds^2}=\frac{(\partial^B\tilde S)(\partial^A\partial_B\tilde S)}{M_a^2}.
\label{twelve}
\eea
\section{Finsler Geometry}
 Finsler geometry provides a geometric background for field theories and a framework to discuss the gravitational dynamics.
 Here we investigate the dynamics and geometry of spinless bosons over a non-metric general length measure Finsler spacetime. Finsler spacetime is actually a natural generalization of the Lorentzian metric theory of gravity. It is possible to extend GR to Finsler geometry if the spacetime is equipped with a general length measure \cite{32}\cite{33} i.e.,

 \bea{}
 g_{\Xi\Sigma}=\frac{1}{2}\frac{\partial^2 F^2(x,y)}{\partial{y}^\Xi\partial{y}^\Sigma}
 \label{aa}
 \eea

where $F$ is the separation between two events on the worldline.
The Einstein-Hilbert action including matter field interaction is given by

\bea{}
S[g,\sigma_i]=\int\limits_
M d^4x\sqrt{|\hat{g}|}\left(\hat{R}+k\hat{L}_M[g,\sigma_i]\right)\nonumber
\eea
Replacing
\bea{}
\sqrt{|\hat{g}|}=\sqrt{|\hat{g}|}\sqrt{|h|}
\label{26}
\eea
\bea{}
\hat{R}=\hat{R}_{\Xi\Sigma}
\label{27}
\eea
The Einstein-Hilbert action could then be rewritten for Finsler space in terms of a homogenous Finsler function $F$ (length between two events on worldline) over a tangent bundle $TM$. Also we consider the circle $S^1_P$ to be fibered over each point of the $2nD$-space manifold $M$ in the tangent space $T_PM$.
\bea{}
S_P^1=\left\{y\in T_PM\mid\sqrt{g_{\mid P}(y,y)}=1\right\}\nonumber
\eea
The Einstein-Hilbert action could then be written as an action on the circle bundle $\sum$ which is subset of tangent bundle $TM$ (obtained by union over all tangent spaces $T_PM$ i.e $TM=\bigcup_{P\in M}T_PM$) as
\bea{}
S_P^1\subset T_PM\nonumber
\eea

Introducing the notion
\bea{}
q^\Sigma=(\hat{x}^\beta,\theta^\alpha),\,\,\,\,\,\,\ \hat{x}^\beta=x^1_1,x_1^2,...,x_n^1,x_n^2 ,\,\,\,\,\,\ \alpha=1,2,...,n \nonumber
\eea
The curvature of such space is specified by $1+3n$ dimensional equation as
\bea{}
P_s\left(\hat{R}_{\Xi\Sigma}+k\hat{L}_M\right)=\hat{R}_{\Xi\Sigma}+k\hat{L}_M
\label{thirteen}
\eea
 In this equation $P_s$ is the operator to preserve symmetry between different particles $x_c^\lambda$ and $x_a^\lambda$, $\hat{R}_{\Xi\Sigma}$ is Ricci tensor, $\hat{L}_M$ is matter lagrangian and $k$ is coupling constant representing the extent of interaction between particles and field.

The metric $\hat{g}$ is transformed conformally, splitting it into conformal function $\sigma(q,t_1)$ and a flat part $\eta$, as
\bea{}
\hat{g}_{\Xi\Sigma}=\sigma^{\frac{4}{3n-1}}\eta_{SA}
\label{fifteen}
\eea

This conformal mapping preserves the local angles and does not change the physics.
The inverse of the metric is given by
\bea{}
\hat{g}^{\Xi\Sigma}=\sigma^{\frac{-4}{3n-1}}\eta^{SA}
\label{sixteen}
\eea

The lower Greek and lower Roman index are identified as $\hat{\partial}_\Sigma=\partial_A$ so that the adjoint derivatives are different in each notation i.e.,
\bea{}
\hat{\partial}^\Sigma&=&g^{\Sigma\Lambda}\hat{\partial}_\Lambda=\sigma^{\frac{-4}{3n-1}}\eta^{AL}\partial_L\nonumber
\\
\hat{\partial}^\Sigma&=&\sigma^{\frac{-4}{3n-1}}\partial^A\nonumber
\\
\hat{\partial}_\Sigma&=&\sigma^{\frac{4}{3n-1}}\partial_A\nonumber
\eea
\\
The resulting Einstein-Hilbert action in Finsler space for circle bundle becomes
\bea{}
S[F,\sigma_i]=\int dt_1\int\limits_\Sigma d^{2n}x d^n\theta\sqrt{|\hat{g}||h|}\left(\hat{R}_{\Xi\Sigma}+k\hat{L}_M[g,\sigma_i]\right)~
\label{28}
\eea
where $F$ is the Finsler function. Note that in Finsler spacetime, the matter-field coupling (given by $k$) is different from Einstein's constant.
\\
Using $\hat{R}_{\Xi\Sigma}=\hat{g}_{\Xi\Sigma}\hat{R}$

, applying conformal transformation 

and choosing $R=0$ for an atlas on this manifold, gives.
\bea{}
S[F,\sigma_i]=\int dt_1\int\limits_\Sigma d^{2n}x d^n\theta\sqrt{|\hat{g}||h|}\sigma^{\frac{6-6n}{3n-1}}\times~~~~~~~~~~~~~~~~~~~~~~~~~~~~\nonumber
\\
\left[\sigma^{3n-5/3n-1}\frac{12n}{1-3n}\partial^A\sigma\partial_A\sigma+k\sigma^2L_M\right]~~~~~~~~~~~~\nonumber
\eea
The matter Lagrangian is given by
\bea{}
L_M=\frac{2(\hat{\partial}^\Sigma \tilde{S}_H)(\hat{\partial}_\Sigma \tilde{S}_H)}{2\hat{M}_G}+\frac{\partial \tilde{S}_H}{\partial t_1}-\hat{M}
\label{ff}
\eea
The equation of motion with this Lagrangian then gives
\bea{}
\sigma^{3n-5/(3n-1)}\frac{12n}{1-3n}\frac{\partial^A\partial_A\sigma}{k\sigma}=\frac{2(\hat{\partial}^\Sigma \tilde{S}_H)(\hat{\partial}_\Sigma \tilde{S}_H)}{2\hat{M}_G}+\frac{\partial \tilde{S}_H}{\partial t_1}\nonumber
\\
-\hat{M}~~~~~~~~~~~~~~~~~~~~~~~~
\label{b}
\eea
With the following matching conditions
\bea{}
k=\sigma^{3n-5/(3n-1)}\frac{12n}{1-3n}.\frac{2\hat{M}_G}{\hbar^2}\nonumber
\eea
\bea{}
\sigma(q,t_1)=R(\vec{x}_a,t_1)\nonumber
\eea
\bea{}
\tilde{S}_H(q,t_1)=\tilde{S}(\vec{x}_a,t_1)\nonumber
\eea
\bea{}
M_a=\hat{M}_G\nonumber
\eea
It is clear that Eq.\eqref{b} is identical to Eq.\eqref{nine}
,where $q$ in these equations represents generalized coordinates $q^\Sigma= (\hat{x}^\beta, \theta^\alpha)$.


\subsection{Geometric dual to second Equation}
Using covariant conservation of stress-energy tensor gives (see appendix C)
\bea{}
\frac{(\hat{\partial}_\Sigma \tilde{S}_H)\hat{\nabla}_\Sigma(\hat{\partial}^\Sigma \tilde{S}_H)}{\hat{M}_G}+\hat{\nabla}_\Sigma\frac{\hat{\partial} \tilde{S}_H}{\partial t_1}=0
\label{rr}
\eea
The Levi-Civita connection here is given by
\bea{}
\hat{\Gamma}^\Lambda_{\Sigma\Xi}=\frac{1}{2}\hat{g}^{\Lambda\Delta}\left(\hat{\partial}_\Sigma \hat{g}_{\Xi\Delta}+\hat{\partial}_\Xi \hat{g}_{\Sigma\Delta}-\hat{\partial}_\Delta \hat{g}_{\Sigma\Xi}\right)
\label{hh}
\eea
Using this connection in Eq.\eqref{rr}, we get (for details see appendix C)
\bea{}
\left(\frac{\partial_A\left[\sigma^2(\partial^A \tilde{S}_H)\right]}{\hat{M}_G}+\frac{\partial}{\partial t_1}(\sigma^2)\right)=0
\label{kk}
\eea
which is dual to Eq.\eqref{ten}.
Furthermore, it is important to stress that the role of $\sigma(q,t_1)$ is two-fold; firstly, it describes the probable location of particle at any particular instant of time, secondly, it represents the conformal function of the theory and accounts for the matter-field interaction.

\subsection{Geometric dual to third Equation:}
Guiding equation is given by
\bea{}
\hat{p}^\Sigma=\hat{\partial}^\Sigma \tilde{S}_H
\label{23}
\eea
which is identical to Eq.\eqref{eleven}
\subsection{Geometric dual to trajectory Equation of motion}

The geometric dual to the equation of motion is obtained as (see appendix D)
\bea{}
\frac{d^2\hat{x}^\Sigma}{d\hat{s}^2}=\frac{(\hat{\partial}^\Xi \tilde{S}_H)\hat{\partial}_\Xi(\hat{\partial}^\Sigma \tilde{S}_H)}{M_G^2}
\label{ee}
\eea
where, $\hat{s}=t_1$ represents the single absolute time coordinate. With the given matching conditions from Eq.\eqref{ee} is identical to Eq.\eqref{twelve}. It is to be noted that

\begin{itemize}
  \item Although the Finsler geometry developed in this section follows from GR, the theory is strictly non-local. The non-locality is encoded by
  \begin{enumerate}
    \item \emph{Eq.\eqref{ee}}: This  equation is just parallel to the Newton's equation of motion and clearly reflects that all the particles taken together completely describe each particle in the system.
    \item \emph{Hamilton-Jacobi equation \eqref{b}}: This equation includes quantum potential which induces a force that acts on all the particles in orbit. It is worth noting that, we considered a system of free bosons but the theory induces a potential $Q$ that interlocks the position of particles such that it impossible to insulate one particle from the rest.
   \item \emph{Extra Dimensions}: We developed the model such that each particle resides in its own frame of reference with two spatial and one polar dimension and all (particles) sharing a universal time coordinate $t_1$. The symmetrization postulate requires all the reference frames to be identical such that changes made in any one of the reference frame demands a simultaneous change in the other frames too and hence induces non-locality in the system.
  \end{enumerate}
  \item The dual to the quantum equations Eqs.\eqref{nine}-\eqref{twelve} is presented in ($1+3n)$ dimensional Finsler geometry. The model introduces a $2nD$-plane as base manifold with unit circle fibered over each point of the plane. This creates as a whole with one universal time coordinate $(1+3n)$ dimensions. One must note that although Finsler geometry is merely Riemannian geometry equipped with homogeneous function $F$, its different from gravity. And one could not retrieve gravity from the Finsler framework (developed in this section) simply by swapping the general length measure with the Lorentzian metric. This is because we are working in the classical limit with a higher dimensional space $(n>4)$ and gravity works with $4$ dimensions only.
  \item One must note that, Eq.\eqref{b} and Eqs.\eqref{kk}-\eqref{ee} are merely geometrical rewriting of the quantum equations Eq.\eqref{nine}-\eqref{twelve}. We defined a set of  matching conditions to connect the quantum equations with those in the Finsler geometry. These matching conditions connect the quantum phase $\tilde{S}$ with the Hamilton principle function $\tilde{S}_H$, the amplitude of pilot wave $R$ with the conformal function of the metric $\sigma$ and the mass $M_a$ with the mass $\hat{M}_G$. The coupling constant defined to establish duality between quantum and geometry is

\bea{}
k=\sigma^{3n-5/(3n-1)}\frac{12n}{1-3n}.\frac{2\hat{M}_G}{\hbar^2}
\label{ll}
\eea

This coupling depends upon the conformal function $\sigma^{3n-5/(3n-1)}$. The coupling constant in the particle action is given by

\bea{}
k=\frac{4\pi G}{c^4 }.\frac{1}{y^\Xi y^\Sigma}
\label{ss}
\eea

Comparing Eq.\eqref{ll} and Eq.\eqref{ss}, we are lead to

\bea{}
\sigma^{\frac{3-5n}{3n-1}}=\frac{4\pi G}{c^4}.\frac{1}{y^\Xi y^\Sigma}.\frac{1-3n}{12n}.\frac{\hbar^2}{2\hat{M}_G}\nonumber
\eea

So Eq.\eqref{ll} becomes

\bea{}
k=\frac{4\pi G}{c^4 y^\Xi y^\Sigma}
\label{nn}
\eea

One can deduce from Eq.\eqref{nn} that $k<0$, which implies that the bosons are weakly coupled to the field.
\end{itemize}

\section{application to funk geometry}

The Finsler distance $F$ and the time-like relative Funk distance $F'$ are related as

\bea{}
F'^2(x,y)=\log F^2(x,y)
\label{oo}
\eea

where $F'^2$ in Funk geometry is the timlike separation between two events. $F^2$ in the above equation given by \cite{8}

\bea{}
F^2(x,y)=\frac{d(x,b(x,y))}{d(y,b(x,y))}
\eea

where $d$ represents the non-reversible, non-symmetric Euclidean distance. So, the time-like relative Funk metric in terms of the genuine Finsler metric in Eq.\eqref{aa} could be written as

\bea{}
\hat{g}'_{\Xi\Sigma}=F^{-2} \hat{g}_{\Xi\Sigma}
\eea

The conformal transformation in Funk geometry is given by

\bea{}
\hat{g}'_{\Xi\Sigma}=F^{-2}\sigma^{\frac{4}{3n-1}}\eta_{SA}
\eea

and the inverse Funk metric is transformed conformally as

\bea{}
\hat{g}'^{\Xi\Sigma}=F^2\sigma^{\frac{-4}{3n-1}}\eta^{SA}
\eea

The particle action in this Euclidean time-like relative Funk geometry is given by
\bea{}
S[F',\sigma_i]=\int dt_1\int\limits_\Sigma d^{2n}x d^n\theta\sqrt{|\hat{g'}||h|}\times~~~~~~~~~~~~~~~~~~~~~~~~~~~~\nonumber
\\
\left[\sigma^{-3n-3/3n-1}\hat{g'}_{\Xi\Sigma}\frac{12n}{1-3n}\partial^A\sigma\partial_A\sigma+k\sigma^{\frac{4}{3n-1}}L_M\right]~~~~~~~~~~~~\nonumber
\eea
The matter Lagrangian $L_M$ is
\bea{}
L_M=\frac{2(\hat{\partial}^\Sigma \tilde{S}_H)(\hat{\partial}_\Sigma \tilde{S}_H)}{2\hat{M}_G}+\frac{\partial \tilde{S}_H}{\partial t_1}-\hat{M}
\label{ff}
\eea
The equation of motion with this Lagrangian then gives
\bea{}
\sigma^{3n-9/(3n-1)}\hat{g'}_{\Xi\Sigma}\frac{12n}{1-3n}\frac{\partial^A\partial_A\sigma}{k\sigma}=\frac{2(\hat{\partial}^\Sigma \tilde{S}_H)(\hat{\partial}_\Sigma \tilde{S}_H)}{2\hat{M}_G}\nonumber
\\
+\frac{\partial \tilde{S}_H}{\partial t_1}-\hat{M}~
\label{ii}
\eea
With the following matching conditions
\bea{}
k=F^{-2}\sigma^{3n-5/(3n-1)}\frac{12n}{1-3n}.\frac{2\hat{M}_G}{\hbar^2}\nonumber
\eea
\bea{}
\sigma(q,t_1)=R(\vec{x}_a,t_1)\nonumber
\eea
\bea{}
\tilde{S}_H(q,t_1)=\tilde{S}(\vec{x}_a,t_1)\nonumber
\eea
\bea{}
M_a=\hat{M}_G\nonumber
\eea
Eq.\eqref{ii} is identical to Eq.\eqref{nine}
\subsection{Geometric dual to second equation}
Again using the conservation of stress-energy tensor together with the Levi-Civita connection (as in the previous section) and follwing the same procedure as mentioned in appendix C, one obtains

\bea{}
\left(\frac{\partial_A\left[\sigma^2(\partial^A \tilde{S}_H)\right]}{\hat{M}_G}+\frac{\partial}{\partial t_1}(\sigma^2)\right)=0
\label{jj}
\eea

It easy to see that, with the defined matching conditions for time-like relative Funk geometry, Eq.\eqref{jj} is identical to Eq.\eqref{ten}

\subsection{Geometric Dual to third equation}
The particles are guided along their trajectory by

\bea{}
\hat{p}^\Sigma=\hat{\partial}^\Sigma \tilde{S}_H
\eea

which, with the defined matching conditions is identical to Eq.\eqref{eleven}

\subsection{Geometric dual to trajectory equation of motion}

The geometric dual to the equation of motion in time-like relative Funk geometry is obtained by following the procedure discussed in appendix D, as

\bea{}
\frac{d^2\hat{x}^\Sigma}{d\hat{s}^2}=\frac{(\hat{\partial}^\Xi \tilde{S}_H)\hat{\partial}_\Xi(\hat{\partial}^\Sigma \tilde{S}_H)}{M_G^2}
\eea
It is important to note that
\begin{itemize}
  \item In Finsler geometry:  Eq.\eqref{ee}, Hamilton-Jacobi equation Eq.\eqref{b} and extra dimensions are responsible for the non-local nature. In time-like relative Funk geometry however, there is an additional parameter encoding non-locality in the theory. Since $F'^2$, the time-like separation is when the event is occuring inside the light cone and this as a result can effect another event. Or we can say there exist causality between the events.
  \item Like the Finsler geometry, it is not possible to restore gravity from the time-like relative Funk geometry because we introduced extra dimensions and an absolute single time coordinate $t_1$ shared by all the bosons.
  \item We specified a set of matching conditions to show that quantum equations Eqs.\eqref{nine}-\eqref{twelve} have duality in time-like relative Funk geometry. These matching conditions connect the quantum phase $\tilde{S}$ with the Hamilton principle function $\tilde{S}_H$, the amplitude of pilot wave $R$ with the conformal function of the metric $\sigma$ and the mass $M_a$ with the mass $\hat{M}_G$. The coupling constant defined in the time-like relative Funk geometry is
\bea{}
k=F^{-2}\sigma^{3n-5/(3n-1)}\frac{12n}{1-3n}.\frac{2\hat{M}_G}{\hbar^2}
\label{qq}
\eea
This coupling depends upon the conformal function $\sigma^{3n-5/(3n-1)}$. The coupling constant in the action is given by
\bea{}
k=\frac{4\pi G}{c^4 }.\frac{1}{y^\Xi y^\Sigma}
\label{mm}
\eea
Comparing Eq.\eqref{qq} and Eq.\eqref{mm} gives

\bea{}
\sigma^{\frac{3n-5}{3n-1}}=\frac{4\pi G}{c^4}.\frac{1}{y^\Xi y^\Sigma}.F^2.\frac{1-3n}{12n}.\frac{\hbar^2}{2\hat{M}_G}\nonumber
\eea

So Eq.\eqref{qq} becomes

\bea{}
k=\frac{4\pi G}{c^4 y^\Xi y^\Sigma}
\label{pp}
\eea

Note from Eq.\eqref{pp} that in a time-like relative Funk geometry also $k<0$ showing weak matter-field coupling. One can see that in this specific framework and with this choice of matching conditions, the two geometries lead to the same result. The only difference is that, in time-like relative Funk geometry the metric is a time-like distance function. We could picture this as if time-like relative Funk space a Finsler space equipped with time-like distance function.
This is similar to as observed in \cite{8} in the framework of convex geometry with time-like norm defined as distance between two points.
\end{itemize}

\section{summary}
We studied a system of $n$ spin-less free bosons in the pilot wave limit, consisting of a set of four quantum equations.
We showed that, when gravity is modified by introducing extra dimensions in the system, the equations following from an extended Hilbert (Finsler) action are dual to the non-relativistic quantum equations in the pilot wave limit.
We also observed that, even though particles are free in this spacetime, a quantum potential $Q$ is induced into the system.
This, along with other parameters (symmetrization postulate, the force $\sum\limits_{c=1}^n(\partial_c^d\tilde S)(\partial_a^b\partial_{cd}\tilde S)/M_a^2$ ), introduces non-locality into the system.
To translate between the quantum description and the geometrical description,
we defined a specific set of matching conditions. We further analyzed the relation between Finsler metric and time-like relative Funk metric given by Eq.\eqref{oo}, and generalized our results to the case of spacetime equipped with $F'^2$. The couplings needed to match the equations in Finsler geometry and timelike relative Funk geometry with
the quantum equations are respectively
\bea{}
k=\sigma^{3n-5/(3n-1)}\frac{12n}{1-3n}.\frac{2\hat{M}_G}{\hbar^2}\nonumber
\eea
\bea{}
k=F^{-2}\sigma^{3n-5/(3n-1)}\frac{12n}{1-3n}.\frac{2\hat{M}_G}{\hbar^2}.\nonumber
\eea
Even though
two couplings seem to be different, one can see from Eq.\eqref{nn} and Eq.\eqref{pp}, that the two couplings are actually equal, thus in both frameworks (Finsler and time-like relative Funk) bosons are weakly coupled to the field.
Finally, in the context of this work we observed that time-like relative Funk space is merely a Finsler space with a time-like distance function.
\section{Appendix}

\subsection{First Equation}
From \eqref{bb}
\bea{}
\left(\sum\limits_{a=1}^n\frac{\hbar}{2M_a}\partial_a^b\partial_{ab}+i\partial_0\right)R(t_1,\vec{x}_a)e^{i\left(\tilde S(t_1,\vec{x}_a)-Mt_1\right)/\hbar}=0~~
\eea
From this equation, we get
\bea{}
\sum\limits_{a=1}^n\frac{\hbar}{2M_a}\partial_a^b\partial_{ab}R(t_1,\vec{x}_a)e^{i\left(\tilde S(t_1,\vec{x}_a)-Mt_1\right)/\hbar}
+\nonumber\\
\sum\limits_{a=1}^n\frac{\hbar}{2M_a}\partial_a^bR(t_1,\vec{x}_a)e^{i\left(\tilde S(t_1,\vec{x}_a)-Mt_1\right)/\hbar}\frac{i}{\hbar}\left(\partial_{ab}\tilde S\right)\nonumber+\\\frac{iR(t_1,\vec{x}_a).i}{\hbar}e^{i\left(\tilde S(t_1,\vec{x}_a)
-Mt_1\right)/\hbar}.\frac{\partial}{\partial t_1}\left(\tilde S-Mt_1\right)=0
\eea
 For large $t_1$, $\frac{\partial R}{\partial t_1}$ on the average is zero. We used the Taylor expansion in the above equation and picked up the real part of the resulting equation to get
\bea{}
\sum\limits_{a=1}^n\frac{\hbar^2}{2M_a}\frac{\partial_a^b\partial_{ab}R(t_1,\vec{x}_a)}{R(t_1,\vec{x}_a)}=\sum\limits_{a=1}^n\frac{(\partial_a^b\tilde S)(\partial_{ab}\tilde S)}{2M_a}\nonumber
\\+\dot{\tilde S}-M
\eea
where, the dot represents the derivative with respect to $t_1$. Using the definition $Q=\sum\limits_{a=1}^n\frac{\hbar^2}{2M_a}\frac{\partial_a^b\partial_{ab}R(t_1,\vec{x}_a)}{R(t_1,\vec{x}_a)}$, we can write
\bea{}
Q=\sum\limits_{a=1}^n\frac{(\partial_a^b\tilde S)(\partial_{ab}\tilde S)}{2M_a}+\dot{\tilde S}-M,
\eea
\subsection{Fourth Equation}
To find the equation of motion, we used the guiding equation $
\frac{dx_a^b}{ds}=\frac{\partial_a^b\tilde S}{M_a}$
This leads to
\bea{}
\frac{d^2x_a^b}{ds^2}=\frac{d}{ds}\left(\frac{\partial_a^b\tilde S}{M_a}\right)
\eea
\bea{}
\Rightarrow \frac{d^2x_a^b}{ds^2}=\sum\limits_{c=1}^n\partial_{cd}\frac{(\partial_c^d\tilde S)(\partial_a^b\tilde S)}{M_a^2}
\eea
\subsection{Geometric dual to second Equation}
The stress-energy tensor is given by
\bea{}
T^{\Sigma\Xi}=-2\frac{\delta L_M}{\delta \hat{g}_{\Sigma\Xi}}+\hat{g}^{\Sigma\Xi}L_M
\eea
The above equation upon substituting matter Lagrangian \eqref{ff} gives
\bea{}
\hat{T}^{\Sigma\Xi}=-2\frac{\delta}{\delta \hat{g}_{\Sigma\Xi}}\left(\hat{g}_{\Sigma\Xi}\frac{(\hat{\partial}^\Sigma \tilde{S}_H)(\hat{\partial}^\Xi \tilde{S}_H)}{\hat{M}_G}+\frac{\hat{\partial} \tilde{S}_H}{\partial t_1}-\hat{M}\right)\nonumber\\+
\frac{\hat{g}^{\Sigma\Xi}(\hat{\partial}^\Sigma\tilde{S}_H)(\hat{\partial}_\Sigma \tilde{S}_H)}{\hat{M}_G}+\hat{g}^{\Sigma\Xi}\frac{\hat{\partial} \tilde{S}_H}{\partial t_1}-\hat{g}^{\Sigma\Xi}\hat{M}~~~
\eea
Differentiating the term inside the bracket with respect to metric, we get
\bea{}
\hat{T}^{\Sigma\Xi}=-\frac{2(\hat{\partial}^\Sigma \tilde{S}_H)(\hat{\partial}^\Xi \tilde{S}_H)}{\hat{M}_G}+\nonumber
\\ \hat{g}^{\Sigma\Xi}\left(\frac{(\hat{\partial}^\Sigma \tilde{S}_H)(\hat{\partial}_\Sigma \tilde{S}_H)}{\hat{M}_G}+
\frac{\hat{\partial} \tilde{S}_H}{\partial t_1}-\hat{M}\right)
\eea
Using covariant conservation of stress-energy tensor $\hat{\nabla}_\Sigma \hat{T}^{\Sigma\Xi}=0$ gives
\bea{}
\hat{\nabla}_\Sigma[\frac{-2(\hat{\partial}^\Sigma \tilde{S}_H)(\hat{\partial}^\Xi \tilde{S}_H)}{\hat{M}_G}
+\nonumber\\
\hat{g}^{\Sigma\Xi}\left(\frac{(\hat{\partial}^\Sigma \tilde{S}_H)(\hat{\partial}_\Sigma\tilde{S}_H)}{\hat{M}_G}+
\frac{\hat{\partial} \tilde{S}_H}{\partial t_1}-\hat{M}\right)]=0
\label{gg}
\eea

Applying the covariant derivative and assuming the metric to be covariantly conserved i.e $\hat{\nabla}_\Sigma \hat{g}^{\Sigma\Xi}=0$, \eqref{gg} simplifies to
\bea{}
-\frac{2(\hat{\partial}^\Xi \tilde{S}_H)\hat{\nabla}_\Sigma(\hat{\partial}^\Sigma \tilde{S}_H)}{\hat{M}_G}-\frac{(\hat{\partial}^\Sigma \tilde{S}_H)\hat{\nabla}^\Xi(\hat{\partial}_\Sigma \tilde{S}_H)}{\hat{M}_G}
\nonumber\\+
\frac{g^{\Sigma\Xi}(\hat{\partial}_\Sigma \tilde{S}_H)\hat{\nabla}_\Sigma(\hat{\partial}^\Sigma \tilde{S}_H)}{\hat{M}_G}+g^{\Sigma\Xi}\hat{\nabla}_\Sigma\frac{\partial \tilde{S}_H}{\partial t_1}=0
\eea
From above equation it follows that
\bea{}
\frac{(\hat{\partial}^\Xi \tilde{S}_H)\hat{\nabla}_\Sigma(\hat{\partial}^\Sigma \tilde{S}_H)}{\hat{M}_G}=0
\eea
\bea{}
\frac{(\hat{\partial}^\Sigma \tilde{S}_H)\hat{\nabla}^\Xi(\hat{\partial}_\Sigma \tilde{S}_H)}{\hat{M}_G}=0
\eea
\bea{}
\frac{(\hat{\partial}_\Sigma \tilde{S}_H)\hat{\nabla}_\Sigma(\hat{\partial}^\Sigma \tilde{S}_H)}{\hat{M}_G}+\hat{\nabla}_\Sigma\frac{\hat{\partial} \tilde{S}_H}{\partial t_1}=0
\label{c}
\eea
Using the Levi-Civita connection \eqref{hh}, \eqref{c} reads
\\
For first term (a part $\hat{\nabla}_\Sigma(\hat{\partial}^\Sigma \tilde{S}_H)$ of it)
\bea{}
\hat{\nabla}_\Sigma(\hat{\partial}^\Sigma \tilde{S}_H)=\hat{\partial}_\Sigma(\hat{\partial}^\Sigma \tilde{S}_H)+
\nonumber\\\frac{1}{2}\hat{g}^{\Lambda\Delta}\left(\hat{\partial}_\Sigma \hat{g}_{\Xi\Delta}+\hat{\partial}_\Xi \hat{g}_{\Sigma\Delta}-\hat{\partial}_\Delta \hat{g}_{\Sigma\Xi}\right)(\hat{\partial}^\Sigma \tilde{S}_H)=0
\eea
Using conformal transformation and doing some mathematics gives
\bea{}
\sigma^{\frac{-2-6n}{3n-1}}\partial_A(\sigma^2\partial^A\tilde{S}_H)=0
\label{d}
\eea
For the second term in \eqref{c}

\bea{}
\hat{\nabla}_\Sigma\frac{\hat{\partial} \tilde{S}_H}{\partial t_1}=\hat{\partial}_\Sigma\frac{\hat{\partial} \tilde{S}_H}{\partial t_1}
\eea
Using conformal transformation, we obtain
\bea{}
\nabla_\Sigma\frac{\hat{\partial} \tilde{S}_H}{\partial t_1}=\frac{\partial}{\partial t_1}(\partial_A \tilde{S}_H)
\label{e}
\eea
\\
From \eqref{c} with first \eqref{d} and second term \eqref{e}
\bea{}
\frac{\sigma^{\frac{-2-6n}{3n-1}}(\hat{\partial}_\Sigma \tilde{S}_H)\partial_A\left[\sigma^2(\partial^A \tilde{S}_H)\right]}{\hat{M}_G}+\frac{\partial}{\partial t_1}(\partial_A \tilde{S}_H)=0
\eea
%
After some manipulations (In doing so we limit the temporal and spatial derivatives up to quadratic terms in $\sigma$ only), we obtain
\bea{}
\left(\frac{\partial_A\left[\sigma^2(\partial^A \tilde{S}_H)\right]}{\hat{M}_G}+\frac{\partial}{\partial t_1}(\sigma^2)\right)=0
\eea
\subsection{Geometric dual to the trajectory Equation of motion}
The total derivative is
\bea{}
\frac{d}{d\hat{s}}&=&\frac{d\hat{x}^\Xi}{d\hat{s}}\hat{\partial}_\Xi
\eea

Applying this relation to momenta, we get


\bea{}
\frac{{d^2\hat{x}^\Sigma}}{d\hat{s^2}}=\hat{\partial}_\Xi\frac{{d\hat{x}^\Xi}}{d\hat{s}}\frac{\hat{\partial}^\Sigma \tilde{S}_H}{\hat{M}_G}
\eea
Using \eqref{23}, this leads to

\bea{}
\frac{d^2\hat{x}^\Sigma}{d\hat{s}^2}=\frac{(\hat{\partial}^\Xi \tilde{S}_H)\hat{\partial}_\Xi(\hat{\partial}^\Sigma \tilde{S}_H)}{M_G^2}
\eea


\begin{thebibliography}{99}

\expandafter\ifx\csname natexlab\endcsname\relax\def\natexlab#1{#1}\fi
\expandafter\ifx\csname bibnamefont\endcsname\relax
  \def\bibnamefont#1{#1}\fi
\expandafter\ifx\csname bibfnamefont\endcsname\relax
  \def\bibfnamefont#1{#1}\fi
\expandafter\ifx\csname citenamefont\endcsname\relax
  \def\citenamefont#1{#1}\fi
\expandafter\ifx\csname url\endcsname\relax
  \def\url#1{\texttt{#1}}\fi
\expandafter\ifx\csname urlprefix\endcsname\relax\def\urlprefix{URL }\fi
\providecommand{\bibinfo}[2]{#2}
\providecommand{\eprint}[2][]{\url{#2}}
\bibitem{1}S. I. Nojiri, and S. D. Odintsov, Mimetic F (R) gravity: inflation, dark energy and bounce. \emph{Mod. Phys. Lett. A}, 29(40) (2014), 1450211.
\bibitem{2}A. G. Riess, \emph{et.al.},  Type Ia Supernova Discoveries at $z > 1$ From the Hubble Space
Telescope: Evidence for Past Deceleration and Constraints on Dark
Energy Evolution. \emph{Astrophys. J}, 607 (2004) 665.
\bibitem{4}A. H. Chamseddine and V. Mukhanov, Mimetic dark matter. \emph{JHEP}, 11 (2013) 135.
\bibitem{5}D. Momeni, R. Myrzakulov and E. G\"{u}dekli, Cosmological viable Mimetic $f(R)$ and $f(R, T)$ theories via Noether
symmetry. \emph{Int. J. Geom. Methods Mod. Phys.}, 12 (2015) 1550101.
\bibitem{6}A. V. Astashenok, S. D. Odintsov and V. K. Oikonomou, Modified Gauss-Bonnet gravity with Lagrange multiplier constraint as mimetic theory. \emph{Classical and Quantum Gravity}, 32 (2015) 185007.
\bibitem{32} C. Pfeifer and M. Wohlfarth, Finsler spacetimes and gravity. In Relativity and Gravitation (pp. 305-308). Springer, Cham (2014)
\bibitem{7}H. Busemann, Time-like spaces. Dissertationes Math. Rozprawy Mat. 53 (1967) 52 pp.
\bibitem{8} A. Papadopoulos, and S. Yamada, Time-like Hilbert and Funk geometries. arXiv preprint arXiv:1602.07072 (2016).
\bibitem{9} I. Tavernelli, On the Geometrization of Quantum Mechanics.  \emph{Ann. Phys}. 371, 239-253 (2016).
\bibitem{12} S. Abraham, P. F. de Cordoba, J. M. Isidro, J. L. G. Santander,  The Ricci flow on Riemann surfaces. arXiv preprint arXiv:0810.2236 (2008).
\bibitem{13} S. Abraham, P. F. de Cordoba, J. M. Isidro, J. L. G. Santander, A mechanics for the Ricci flow. \emph{Int. J.
Geom. Methods Mod. Phys}. 6, 759-767 (2009).
\bibitem{14} B. Koch, Relativistic Bohmian mechanics from scalar gravity. arXiv:0810.2786 [hep-th] (2008).
\bibitem{15} B. Koch, Quantizing geometry or geometrizing the quantum?. \emph{AIP Conf. Proc.} 1232, 313 (2010).
1063/1.3431507. arXiv:1004.2879 [hep-th]
\bibitem{16} B. Koch, Geometrizing the Quantum - A Toy Model. \emph{ AIP Conf. Proc}. 1196, 161 (2009).
1063/1.3284379. arXiv:1004.3240 [gr-qc]

\bibitem{18}B. Koch and N. Rojas, An Angular Formalism for Spin One Half. \emph{Int. J. Geom. Methods Mod. Phys}. 11,
1450029 (2014).
\bibitem{19}D. Acosta, P. Fernandez de Cordoba, J. M. Isidro, J. L. G. Santander,
 An entropic picture of emergent quantum mechanics. \emph{Int. J. Geom. Methods Mod. Phys}. 9, 1250048 (2012).
\bibitem{20} S. H. Mehdipour, Entropic force approach to noncommutative Schwarzschild black holes signals a failure of current physical ideas. \emph{Eur. Phys. J. Plus} 127, 80 (2012).
\bibitem{21} T. S. Bir, P. Vn, Splitting the source term for the Einstein equation to classical and quantum parts. \emph{Found. Phys.} 45(11), 1465 (2015).
\bibitem{22} H. Nikolic, Bohmian particle trajectories in relativistic bosonic
quantum field theory. \emph{Found. Phys. Lett}. 17, 363 (2004)
\bibitem{23} B. Koch, Higher-dimensional geometric description of the quantum Klein-Gordon equation. \emph{Int. J.Geom. Methods Mod. Phys.} 10(9), 1320014 (2013).
\bibitem{top1} E. Witten, Constraints on supersymmetry breaking. \emph{Nucl. Phys. B} 202(2), 253-316 (1982).
\bibitem{top2} L. \'{A}lvarez-Gaum\'{e}, A note on the Atiyah-Singer index theorem. \emph{J. Phys. A Math. Gen.} 16(18), 4177-4182 (1983).
\bibitem{top3} L. \'{A}lvarez-Gaum\'{e}, E. Witten, Gravitational anomalies. \emph{Nucl. Phys. B} 234(2), 269-330 (1984).
\bibitem{top4} J.P. Gauntlett, Low-energy dynamics of  $\mathcal{N}=2$ supersymmetric monopoles. \emph{Nucl. Phys. B} 411, 443-460 (1994).
\bibitem{top5} M. A. Wasay, Supersymmetric quantum mechanics and topology. \emph{Adv. High Energy Phys.} 2016, 3906746 (2016).
\bibitem{top6} T.J. Hollowood, T. Kingaby, A comment on the $\chi_y$ genus and supersymmetric quantum mechanics. \emph{Phys. Lett. B} 566, 258-262 (2003).
\bibitem{26} G. Bertoldi, A.E. Faraggi, M. Matone, Equivalence principle,
higher dimensional Mobius group and the hidden antisymmetric
tensor of Quantum Mechanics. \emph{Class. Quantum Gravity} 17, 3965
(2000).
\bibitem{27}D. Bohm, A suggested interpretation of the quantum theory in terms of" hidden" variables. I.\emph{ Phys. Rev}. 85, 166 (1952)
\bibitem{28}D. Bohm, A Suggested Interpretation of the Quantum Theory in Terms of "Hidden" Variables. II. \emph{Phys. Rev}. 85, 180 (1951)
\bibitem{29}M. A. Wasay, A. Bashir, B. Koch, A. Ghaffar, Geometric description of the Schrödinger equation in $(3n+ 1)$-dimensional configuration space. \emph{ Int. J. Geom.
Methods Mod. Phys.} 14, 1750149 (2017).
\bibitem{30}M. A. Wasay, A. Bashir, Two particle entanglement and its geometric duals.\emph{ Eur. Phys. J. C.} 820 (2017)


\bibitem{33} C. Pfeifer, M. N. Wohlfarth, Finsler geometric extension of Einstein gravity. \emph{Phys. Rev. D} 85(6), 064009 (2012)
\bibitem{34} C. L\"{a}mmerzahl, V. Perlick, Finsler geometry as a model for relativistic gravity. \emph{Int. J. Geom. Methods Mod. Phys}. 15(supp01), 1850166 (2018)
\bibitem{35} H. Rund, The differential geometry of Finsler spaces (Vol. 101). Springer Science and Business Media (2012)
\bibitem{36} C. Pfeifer, The Finsler spacetime framework. Backgrounds for physics beyond metric geometry (2013).
\end{thebibliography}
\end{document}